\documentclass[journal,comsoc]{IEEEtran}
\usepackage[utf8]{inputenc}
\usepackage[T1]{fontenc}
\usepackage{url}
\usepackage{ifthen}
\usepackage[hypertexnames=false]{hyperref}
\usepackage[cmex10]{amsmath}
\usepackage{enumerate}
\usepackage{amssymb}
\usepackage{algorithm}
\usepackage{algorithmicx}
\usepackage{algpseudocode}
\usepackage{hhline}
\usepackage{mathrsfs}
\usepackage{bm}
\usepackage{tikz}
\usetikzlibrary{arrows}
\usepackage[caption=false]{subfig}
\usepackage{graphicx,booktabs,multirow}
\usepackage{epstopdf}
\usepackage{tabularx}
\usepackage{cite}
\usepackage{xcolor}
\ifodd 1

\newcommand{\com}[1]{{\color{blue}#1}}
\else

\newcommand{\com}[1]{}
\fi

\begin{document}

\title{Wireless Personal Agent: Extending Wireless Intelligence from Networks to Terminals}

\author{Jiedan Tan, Fang Liu,~\IEEEmembership{Member,~IEEE},  Jingwen Tong,~\IEEEmembership{Member,~IEEE}, Shengli Zhang~\IEEEmembership{Senior Member,~IEEE}, \\ Jun Zhang,~\IEEEmembership{Fellow,~IEEE}, and Wing Shing Wong,~\IEEEmembership{Life Fellow,~IEEE}
\thanks{Jiedan Tan, Fang Liu, Jingwen Tong, and Shengli Zhang are with the College of Electronics and Information Engineering, Shenzhen University, Shenzhen, China (e-mails: 2510044003@mails.szu.edu.cn; \{liuf, eejwentong, zsl\}@szu.edu.cn;
Jun Zhang is with the Department of Electronic and Computer Engineering, The Hong Kong University of Science and Technology, Hong Kong, SAR (e-mail: eejzhang@ust.hk).
Wing Shing Wong is with the Department of Information Engineering,
The Chinese University of Hong Kong (CUHK), Hong Kong, SAR (e-mail:
wswong@ie.cuhk.edu.hk).}
}


\maketitle

\begin{abstract}
Wireless networks are evolving from connectivity-oriented infrastructures into intelligent and personalized service platforms. Existing wireless intelligence remains centered on network-side optimization, improving objectives such as throughput, latency, and coverage. Nevertheless, besides network performance, wireless intelligence also depends on user-perceived experience via application context, mobility routine, service cost, privacy preference, and long-term usage behavior. This article proposes \emph{WISPA}, a \emph{Wireless Intelligent Self-evolving Personal Agent} framework for automated terminal-side resource management based on large language model (LLM)-based agent. To overcome the resource constraints on terminals, WISPA decouples the latency-sensitive online resource execution from offline LLM agent reflection. In this way, a lightweight online executor makes deterministic resource decisions using interpretable preference parameters; While an offline LLM agent analyzes terminal-side traces, refines user profiles, and updates online preference parameters for subsequent decisions. At last, we demonstrate the practical applicability and benefits of WISPA for terminal-side resource allocations on a campus commute route. Numerical results show that WISPA learns user-specific connection styles and adapts access decisions as preferences change.

\end{abstract}


\section{Introduction}
Wireless networks are evolving from connectivity-oriented infrastructures into intelligent and personalized service platforms~\cite{liang2026large}. Emerging services, such as immersive media, mobile AI assistants, interactive learning, and personal productivity applications, require connectivity aligned with application states, mobility patterns, device constraints, and user preferences. Wireless intelligence should therefore consider not only throughput, latency, coverage, and spectral efficiency, but also the experience perceived at the terminal side.

Most existing studies remain network centered~\cite{wang2026empowering}. Base stations, edge platforms, and core networks improve system-level efficiency through planning, spectrum management, scheduling, and operation automation. Recent large language model (LLM)-enabled studies further support telecom knowledge understanding, standards retrieval, wireless reasoning, and network automation~\cite{bariah2024large, bornea2024telco}. However, network-side intelligence alone cannot close the loop of personalized wireless experience, because key quality of experience (QoE) factors, such as application quality, battery burden, service cost, privacy requirements, and usage routines, are observed and acted upon at terminals~\cite{3gpp36214}. A terminal that learns recurrent degradation along a user's route, for example, can adjust access choices before service interruption becomes severe.

Terminal-side resource management therefore complements network-side optimization. Modern terminals have stronger processors, AI accelerators, abundant memory, and storage, and they coordinate heterogeneous resources such as WiFi and cellular interfaces, multiple subscriber identity module (SIM) profiles, connection modes, and privacy-related choices~\cite{itutP10G100}. While the network handles real-time scheduling and system-level efficiency, the terminal can adapt slower preference-related decisions according to individual experience.

This slower-timescale role still imposes strict design constraints. Active-service decisions must remain timely, stable, energy efficient, and interpretable, whereas the underlying preferences vary across applications, locations, and time periods. These preferences are best captured as low-dimensional parameters that balance service-side QoE against terminal-side burdens such as cost, energy, and privacy exposure. Such parameters can be bounded during online execution yet revised from completed usage experience, which makes direct large-model control of real-time decisions both unnecessary and impractical. This separation explains why LLM agents are better suited to reflection than to control: their memory, reasoning, and explanation abilities can interpret heterogeneous traces and justify preference updates offline~\cite{wang2024survey, jiang2026large}, complementing rather than replacing classical optimization, learning-based, and rule-based methods.

\begin{figure*}[!t]
    \centering
    \includegraphics[width=0.8\textwidth]{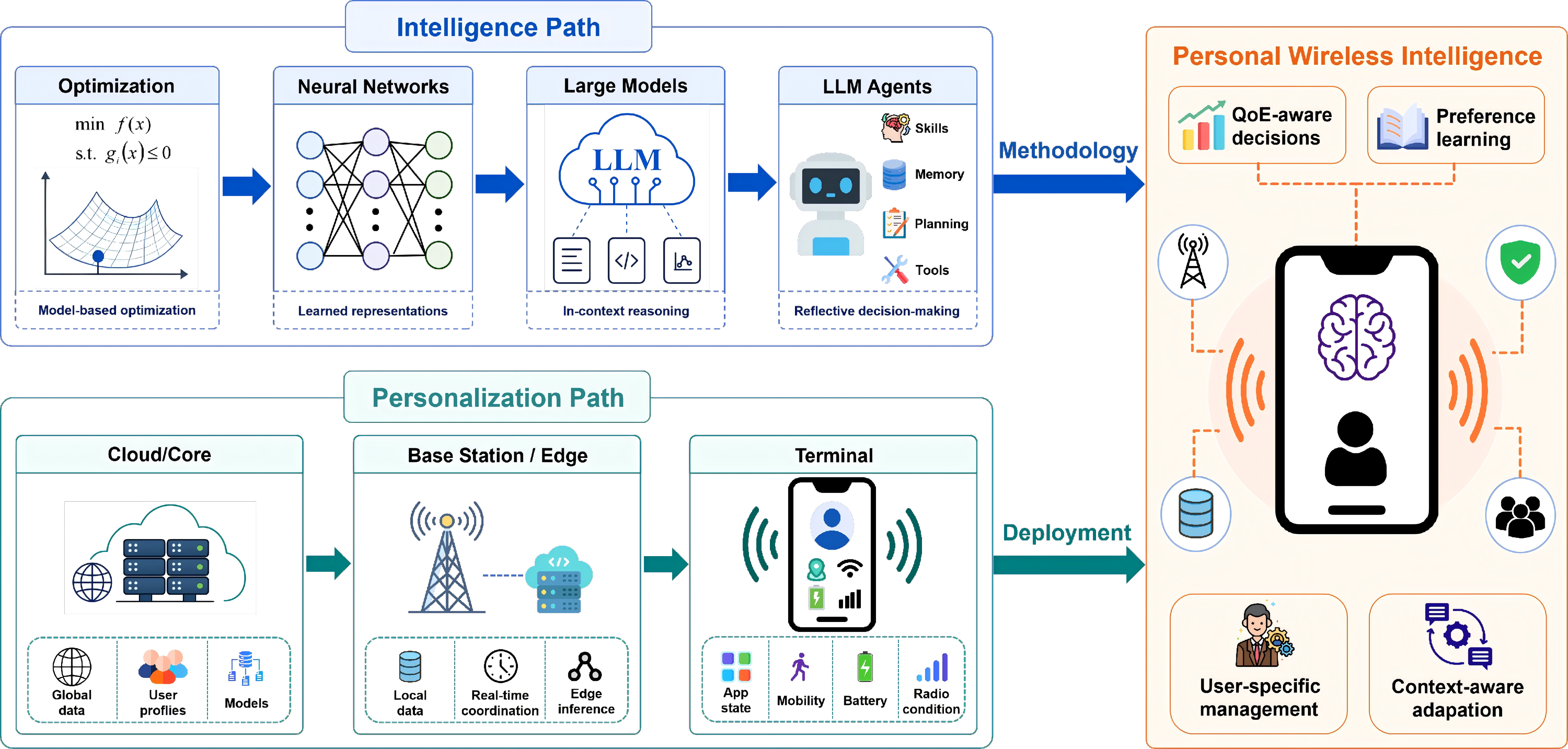}
    \caption{An illustration of the two evolution paths toward personal wireless intelligence. The intelligence path evolves from optimization-driven methods to neural-network-based learning, large-model-based reasoning, and LLM-agent-based intelligence. The personalization path moves intelligence from cloud/core networks to base-station/edge platforms and finally to user terminals. Their intersection leads to terminal-side personal wireless intelligence.}
    \label{fig:framework}
\end{figure*}
Following this principle, we propose \emph{WISPA}, a \emph{Wireless Intelligent Self-evolving Personal Agent} framework for automated terminal-side resource management with LLM Agents. WISPA consists of the online execution stage and the offline reflection stage. At the first stage, an online executor evaluates candidate resource options with the determinate preference parameters and takes the corresponding actions. On the offline stage, an LLM agent analyzes terminal traces, refine user profiles within reflective memory, and updates online preference parameters. If reflection is unavailable or invalid, the terminal keeps the previous parameters. In this way, WISPA treats LLM agents as cognitive engines for long-term personalization rather than as real-time wireless controllers. We demonstrate the practical applicability of WISPA for automated terminal-side resource management on a campus commute rout scenario. Numerical results show that WISPA improves the mean user experience by up to $13.6\%$ over a preference-agnostic online executor. Moreover, WISPA can quickly learn its connection style when user preferences drift.

In summary, this article makes three contributions. First, we articulates a terminal-side paradigm that extends wireless intelligence from networks to terminals. Second, we proposes WISPA as an online/offline LLM-agent framework for automated terminal-side resource management. Third, we presents a proof-of-concept case study for WISP on a campus commute rout scenario.

\section{Toward Personal Wireless Intelligence: Two Evolution Paths}
Wireless intelligence can be understood through two evolution paths: \emph{how} wireless systems become more intelligent, and \emph{where} this intelligence is placed relative to individual users. As Fig.~\ref{fig:framework} illustrates, their intersection points to terminal-side personal wireless intelligence.

\subsection{The Intelligence Path: From Optimization to AI Agents}
Wireless intelligence has progressed through four stages. It began as optimization driven~\cite{lin2006tutorial}, formulating resource allocation, power control, and scheduling as mathematical problems with explicit objectives; this approach offers interpretability but relies on simplified models that weaken under highly dynamic environments and preferences. Neural-network methods then shifted the field toward data-driven learning~\cite{chen2019artificial}, approximating radio propagation, traffic, and mobility for tasks such as channel estimation and beam management; yet most models remain task-specific and lack semantic understanding. Large models added knowledge-aware reasoning and intent understanding~\cite{liu2024llm4cp}, since wireless systems are also knowledge-intensive systems shaped by standards and operational experience; still, their inference latency and hallucination make them ill-suited to direct real-time control.

LLM agents represent the next step~\cite{jingwen2026wirelessagent}. Besides placing a large model in the real-time loop, agents organize reusable \emph{skills}, such as diagnosing service degradation, comparing access options, or revising preference parameters, behind controlled execution \emph{harnesses} that govern data access, tool use, and outputs. This organization lets an agent retain experience, interpret delayed outcomes, and revise future strategies as evidence accumulates, which is precisely what personal wireless intelligence needs to improve across repeated usage periods.

\subsection{The Personalization Path: From Cloud/Core to Terminals}
The second path concerns the location of intelligence. Early network intelligence resided in cloud platforms and core networks, which offer abundant computing and a global view suited to traffic analysis and cross-domain orchestration, but remain far from the user and incur high overhead and latency. Intelligence then moved toward base stations and edge platforms~\cite{mao2017survey}, reducing the distance to radio resources and enabling local scheduling, caching, and near-real-time control; however, edge utilities are organized around cells, slices, and service classes, and are largely agnostic to user-perceived experience that depends on terminal-side factors such as application context, mobility routine, cost, and privacy.

The terminal closes this gap by bringing the control loop to where user context is directly observed and resource actions are directly executed. It can sense application states, mobility routines, and local radio conditions, and execute access selection and connection-mode adjustment. This makes the terminal the natural location for personalized resource management, especially when experience depends on local constraints that are hard to expose to the network. Combined with long-term preference learning, the terminal can develop a user-specific management style that trades service quality against energy, cost, and privacy. 

The convergence of these two paths is enabled by two simultaneous advances. On the one hand, terminals now provide stronger processors, AI accelerators, abundant memory, and multi-interface communication capabilities, allowing them to observe local context and execute resource actions without relying entirely on network-side control. On the other hand, AI has evolved from task-specific models toward agents with memory, reasoning, and tool-use abilities, making it possible to interpret heterogeneous terminal traces and learn from repeated usage experience. Their intersection leads to \emph{Personal Wireless Intelligence}: a self-evolving terminal-side paradigm for wireless resource allocation that builds and refines user profiles, protects private terminal data, and serves as a bridge between user intent and network resources. Rather than optimizing only generic network indicators, personal wireless intelligence translates individual preferences and contextual evidence into adaptive, verifiable, and privacy-preserving wireless decisions, thereby moving wireless intelligence closer to the user. The next section presents WISPA as an architecture that operationalizes this paradigm through lightweight online execution and offline agent-based reflection.

\begin{figure}[!t]
    \centering
    \includegraphics[width=0.48\textwidth]{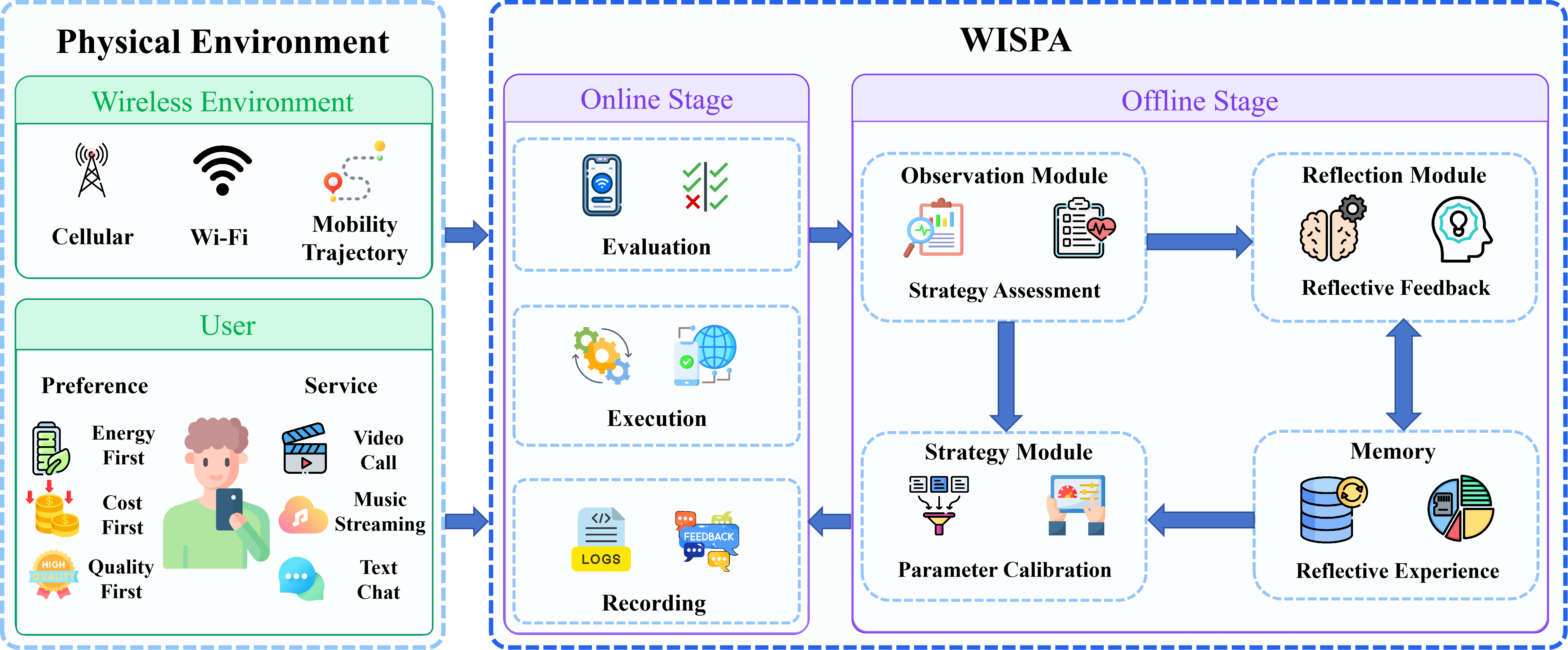}
    \caption{System overview of WISPA. The physical environment includes cellular coverage, campus WiFi, and mobility context, as well as the user's service demand and preference. The online stage performs real-time option execution and records complete usage outcomes, while the offline stage uses role-specialized LLM modules for observation, reflection, and strategy revision, with memory preserving reusable experience for preference calibration.}
    \label{fig:wispa_framework}
\end{figure}
\section{The WISPA Framework}
This section introduces \emph{WISPA, a Wireless Intelligent Self-evolving Personal Agent}, as illustrated in Fig.~\ref{fig:wispa_framework}. WISPA is intended as a feasible terminal-side architecture that makes progress toward personal wireless intelligence under practical constraints: real-time actions must be lightweight and reliable, user data should remain under terminal-side control, and large-model reasoning should improve future decisions without directly manipulating wireless interfaces. To this end, WISPA separates \emph{online execution} from \emph{offline reflection}. The online stage performs deterministic option selection with the current user profile, while the offline stage uses role-specialized LLM modules to interpret completed usage traces, refine user profile, and update the preference parameters for subsequent service periods.

\subsection{Problem Statement}
Consider a user carrying a smartphone through repeated daily service scenarios, such as commuting across a campus, working in different indoor locations, or joining a video call while moving. The user expects stable service, low cost, acceptable energy consumption, and proper privacy protection, but the desired tradeoff among these factors is personal and may change with service type and context. The terminal is the device that directly observes this situation: it senses the active application, radio measurements, battery state, mobility condition, location-related routines, and feasible resource options, e.g., WiFi, cellular profiles, connection modes, or privacy-related configurations. In WISPA, the terminal acts as the user's local wireless agent. It maintains a user profile composed of a bounded low-dimensional \emph{preference parameter} and a reflective memory. The preference parameter converts the user's current tradeoff among service QoE, energy burden, monetary cost, and privacy exposure into an executable scoring rule, while the memory stores longer-term lessons extracted from previous service periods.

The operational task is therefore not to solve the entire network resource allocation problem at the terminal. Instead, WISPA focuses on a narrower but deployable loop: given the current user profile, select a feasible terminal-side resource option during active service, record the outcome, and revise the profile after the service period. For each candidate option, the online executor estimates a service-sensitive QoE term and a set of terminal-side burden terms, then combines them into an \emph{experience score} under the current preference parameter. The selected option is the one with the highest valid score, subject to service-floor and safety checks. Afterward, the offline agent updates the preference parameter and memory from completed traces. This formulation keeps real-time control simple, while leaving personalization to a slower learning process.

\subsection{Online Stage}
The online stage is the real-time execution layer deployed on the terminal. It consists of three lightweight components. First, a context adapter collects service demand, radio measurements, battery state, mobility information, and admissible actions from physical environment. An option evaluator then predicts the QoE and terminal-side burden of each feasible option using simple models, lookup tables, or locally cached estimators. Second, a decision executor applies the preference-parameterized scoring rule and triggers the selected action through verified system interfaces. Third, a record buffer stores the context, candidate scores, selected action, realized service outcome, and representative low-experience events.

This online design avoids LLM inference. For a fixed preference parameter, the decision reduces to structured evaluation and comparison, which makes the behavior fast, reproducible, and interpretable. Service-floor checks prevent low-cost or privacy-friendly options from being selected when they would cause severe service degradation. If required measurements are missing, a candidate action is unsupported, or the scoring result violates a safety rule, the executor falls back to the previous valid policy or a conservative default. 

\subsection{Offline Stage}
The offline stage performs preference calibration during an idle window, such as at the night. Its input is not the raw full-resolution terminal trace by default, but a bounded service records from the online stage. This record include summary statistics, selected representative events, and privacy-filtered context fields. The LLM agent then operates behind a harness that fixes accessible data, required output format, and validation checks. In this way, language reasoning is used to improve future strategies, while the terminal retains control over data exposure and executable actions.

WISPA organizes the offline stage as three role-specialized modules: \emph{observe}, \emph{reflect}, and \emph{revise}. The Observation Module produces a factual diagnosis, the Reflection Module updates reusable user knowledge, and the Strategy Module converts that knowledge into a bounded parameter revision. Keeping these products separate makes the pipeline easier to inspect, debug, and audit than a single unconstrained LLM invocation.

\subsubsection{Observation Module}
The Observation Module is the diagnostic entry of the pipeline. It converts the service records into a round-level assessment of the current strategy: how experience scores evolve over the period, how access behavior changes with service demand and wireless condition, and which low-experience cases dominate the user's dissatisfaction. For example, it can distinguish whether a poor video-call segment is caused by insufficient throughput, excessive delay, repeated handover, high cost, or an overly conservative privacy setting. Its output is a structured assessment, rather than a free-form suggestion, so that later modules can trace each recommendation back to concrete evidence.

\subsubsection{Reflection Module and Memory}
The Reflection Module turns the assessment into reusable user knowledge. Given the current assessment and prior \emph{Memory}, it identifies stable preference cues, temporary anomalies, and recurring context-action patterns. For instance, it may record that the user tolerates moderate quality loss during music streaming to reduce cost, but prefers high-assurance access during video calls near weak WiFi regions. Memory is maintained as a compact profile rather than an unrestricted transcript, which reduces privacy exposure and prevents the agent from overfitting to a single service period. Through this memory, WISPA gradually constructs a terminal-side user profile while keeping the rawest personal traces local.

\subsubsection{Strategy Module}
The Strategy Module turns offline understanding into an executable update. Weighing the latest assessment against the accumulated memory, it proposes how the preference parameter should move before the next service period. This proposal must be structured, bounded, and validated by the harness. Each update is limited to an admissible range and a maximum step size, and it can be checked by a simulator or rule-based validator before deployment. If the recommendation is missing, inconsistent, or unsafe, WISPA keeps the previous parameter as a safe default. The accepted parameter is then loaded by the online executor, closing the online--offline loop. The case study in Section~IV evaluates this feasible loop under heterogeneous access selection and dynamic preference drift.

\section{Case Study and Experimental Results}

This section provides a proof-of-concept case study and conducts several simulations to evaluate the WISPA framework. We consider a campus commute route where a graduate student repeatedly walks from the laboratory to the dormitory while using text interaction, music streaming, and video calling on a smartphone. Along the route, the terminal observes changing WiFi and cellular conditions and can choose among WiFi, an ordinary cellular profile, and a higher-assurance cellular profile. WISPA aims to learn the user's connection style over repeated commutes, so that later online decisions better match the user's tradeoff among service quality, cost, energy burden, and privacy-related preference. 

\begin{figure*}[!t]
    \centering
    \subfloat[Campus Route Map]{%
        \includegraphics[width=0.46\textwidth]{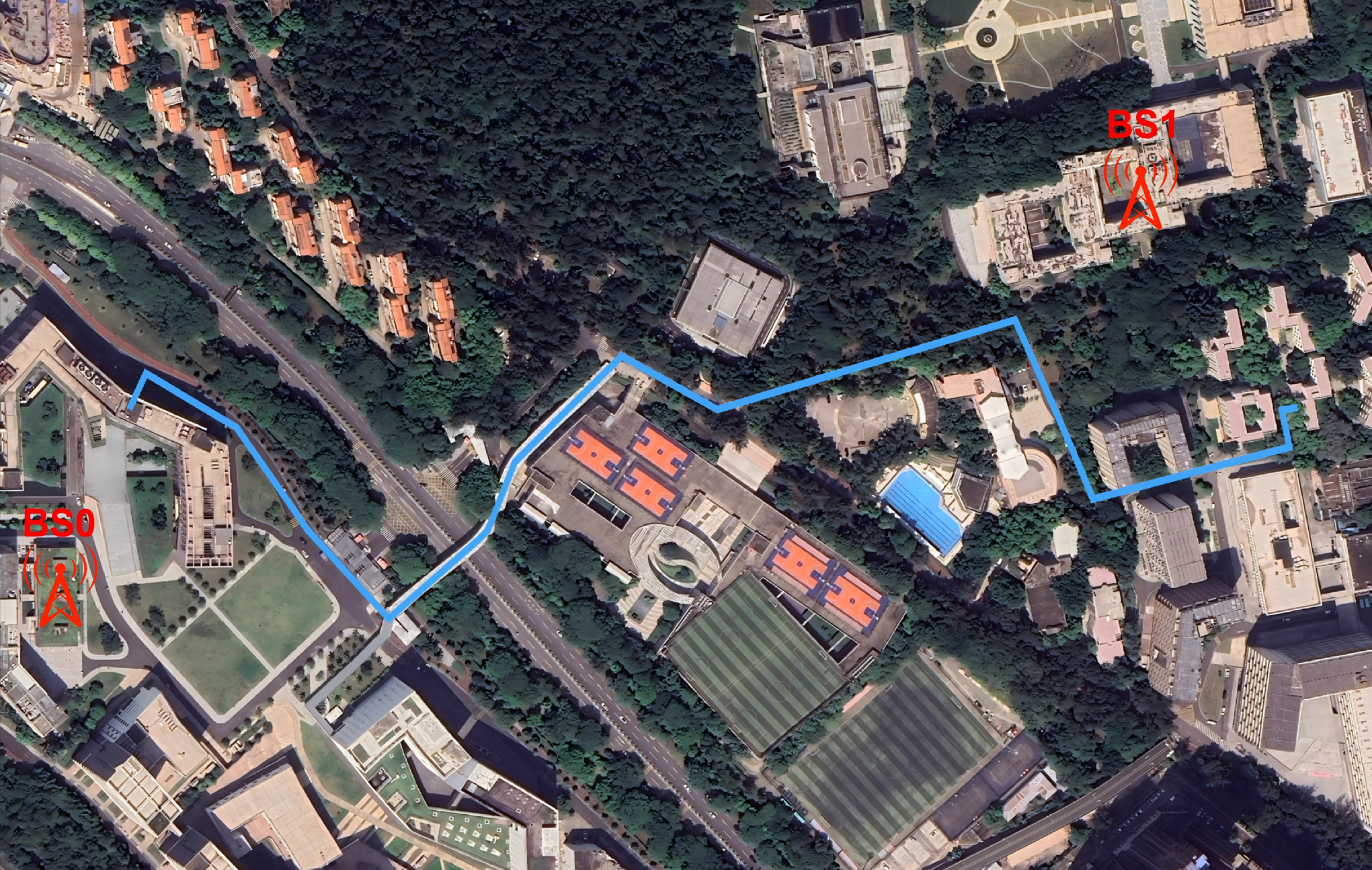}
        \label{fig:case_scene_geometry}}
    \hfill
    \subfloat[Path-Level Cellular Trace]{%
        \includegraphics[width=0.50\textwidth]{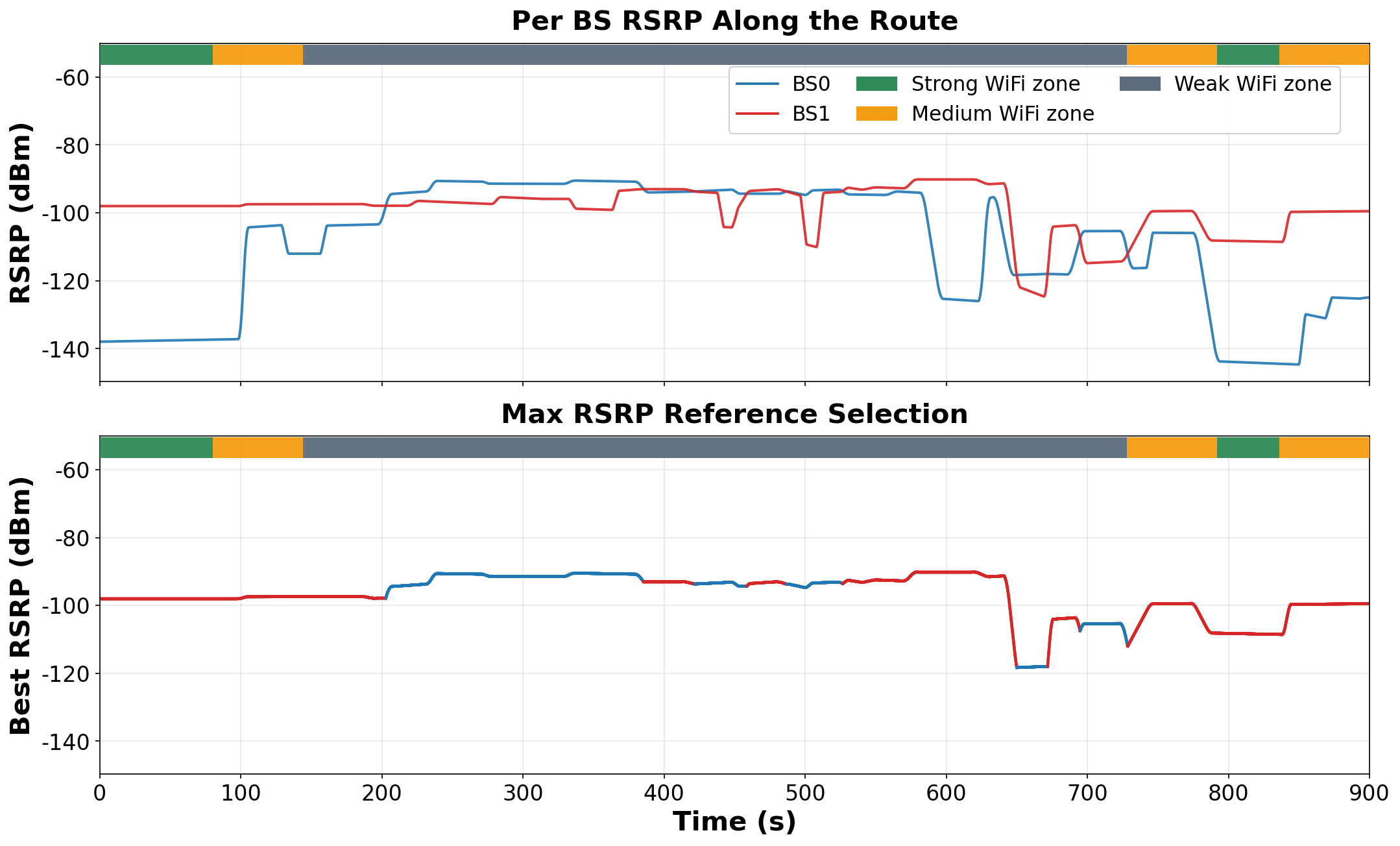}
        \label{fig:case_scene_trace}}
    \caption{Campus route map and path-level wireless trace. The left figure shows the Shenzhen University route, major building footprints, and two cellular base stations. The right figure shows the corresponding path-level cellular trace together with the WiFi strong/medium/weak coverage partition used by the terminal-side agent.}
    \label{fig:case_scene}
\end{figure*}

\subsection{Simulation Setting}
\subsubsection{Scenario}
As shown in Fig.~\ref{fig:case_scene_geometry}, the campus commute route is built from the Shenzhen University layout, with major building footprints and  two cellular base stations placed at distinct locations. The scene spans roughly $900\,\mathrm{m}\times 525\,\mathrm{m}$, contains 35 buildings, and has major building heights between 10 and 40 meters. During one commute, the user moves through open areas, building-edge regions, and dense dormitory-side structures. The terminal experiences vary with changing radio conditions and services.

The cellular trace is generated by ray-tracing over the OpenStreetMap-based geometry, yielding a path-level received-signal pattern along the commute route. We parameterize the trip as a 900 s walking episode and sample the reference signal received power (RSRP) trace every 0.2 s, producing a time-aligned cellular measurement sequence. The RSRP is used as the cellular-side measurement anchor~\cite{3gpp36214}. Fig.~\ref{fig:case_scene_trace} shows the trace together with the WiFi strong/medium/weak coverage partition: strong regions appear near the laboratory and dormitory, medium regions around their edges, and the corridor between them is treated as weak WiFi coverage.

\subsubsection{Unified Wireless Service Modeling} 

We use the \emph{experience score} to quantify performance. The unified service model converts heterogeneous access options, service requirements, and user-side burdens into one comparable score for online selection. It follows four connected stages: (a) describing each candidate access option by a capability profile, (b) mapping the profile and wireless state to quality of service (QoS), (c) interpreting QoS through the active service profile to obtain service-sensitive QoE, and (d) applying a service floor before combining QoE, friendliness score, and preference parameter into the final experience score. The friendliness score captures the burden-side advantage of a candidate; in this experiment, a higher value denotes a lower fee burden.

\paragraph{Candidate capability profiles}

The first stage parameterizes the three access actions introduced above: WiFi, SIM1, and SIM2. Each candidate is represented by a capability profile with four fields: capacity, latency, reliability, and friendliness score. The first three fields are translated into QoS outcomes in the next stage, while the friendliness score is preserved as the burden-side term used in the final experience calculation.

WiFi is region-dependent: strong, medium, and weak regions provide 1.1--2.2 Mb/s with 8--22 ms delay, 280--360 kb/s with 150--220 ms delay, and 178--256 kb/s with 260--460 ms delay, respectively, with reliability tied to the coverage region. Its friendliness score is 1.00, corresponding to the lowest fee burden. SIM1 represents ordinary route-wide cellular access with 350 kHz capacity, 35 ms latency, a $-0.5$ dB reliability offset, and a 0.65 friendliness score. SIM2 represents a higher-assurance cellular option with 1500 kHz capacity, 5 ms latency, a $+7.5$ dB reliability offset, and a 0.25 friendliness score.

\paragraph{Capability-to-QoS mapping}
The second stage converts candidate capabilities into measurable service conditions: throughput, delay, and packet loss. WiFi obtains its QoS conditions directly from the route-aligned coverage regions. For SIM1 and SIM2, the path-level RSRP trace provides the link-quality anchor, while candidate capacity, latency, and reliability offset determine the attainable throughput, base delay, and packet-loss condition. Throughput is further shaped by link quality and loss, and delay increases when loss pressure or throughput shortfall becomes significant.

\paragraph{Service-sensitive QoE}

The third stage accounts for service-dependent experience: the same QoS condition may be acceptable for text interaction but insufficient for video calling. We include three representative services along the route: text interaction, music streaming, and video calling. With reference to standardized 5G QoS Identifier (5QI) service characteristics \cite{3gpp23501}, text interaction is assigned a 300 ms delay target, 20 kb/s throughput floor, and $10^{-6}$ loss tolerance; music streaming uses 250 ms, 320 kb/s, and $10^{-2}$; and video calling uses 120 ms, 1.8 Mb/s, and $10^{-3}$, respectively. QoE is then computed by mapping delay, loss, and throughput into service-sensitive satisfaction terms. Delay and loss follow bounded exponential degradation, while throughput follows a logarithmic saturating response. Thus, text interaction is mainly delay-sensitive, music streaming emphasizes sustained throughput and information-loss control, and video calling stresses low delay, sufficient throughput, and reliable delivery.

\paragraph{Service floor and experience score}

The fourth stage turns service-sensitive QoE into the final decision metric. Before preference is applied, each candidate must satisfy the service floor of the active service. Since terminal services differ in their tolerance to delay, throughput shortage, and information loss \cite{itutG1010}, this floor combines per-dimension severe-violation checks with an aggregate deficit check. If both checks indicate unacceptable degradation, the instantaneous experience score is set to zero, preventing friendliness from compensating for a severe service failure.

For service-acceptable candidates, the final experience score combines QoE and friendliness score through the preference parameter, matching the burden-quality tradeoff defined in Section~III. A usage--revision cycle corresponds to one service period of online execution followed by one offline calibration. The online executor selects the candidate with the highest estimated experience score, and the mean experience score averages the deployed decisions over the complete usage period.

\begin{figure}[!t]
    \centering
    \includegraphics[width=0.45\textwidth]{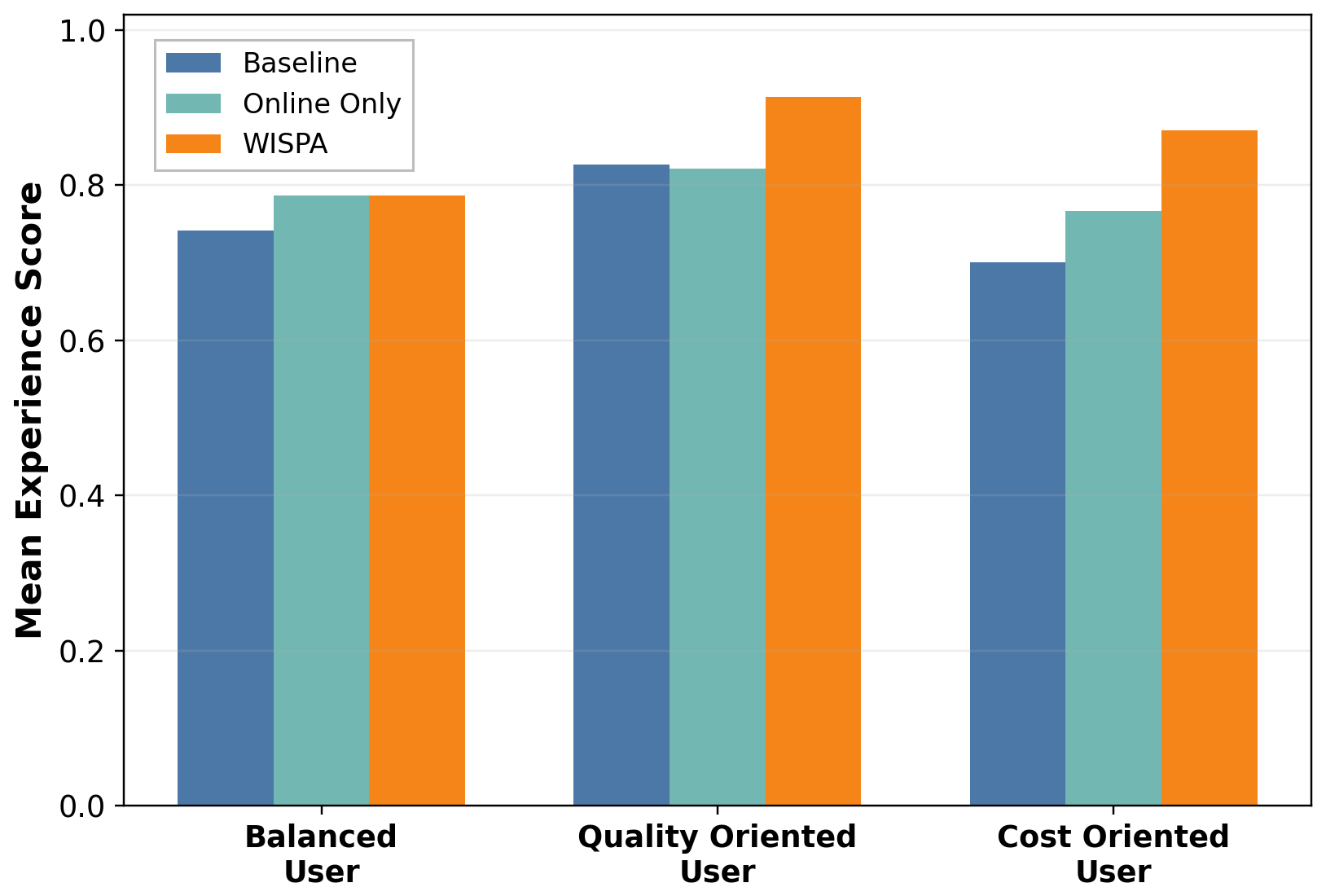}
    \caption{Overall comparison of mean experience score for the Cost Oriented, Quality Oriented, and Balanced users under Baseline, Online Only, and WISPA, using the final strategy obtained after 10 usage--revision cycles.}
    \label{fig:bar}
\end{figure}

\begin{figure*}[!t]
    \centering
    \subfloat[Cost Oriented User]{%
        \includegraphics[width=0.49\textwidth]{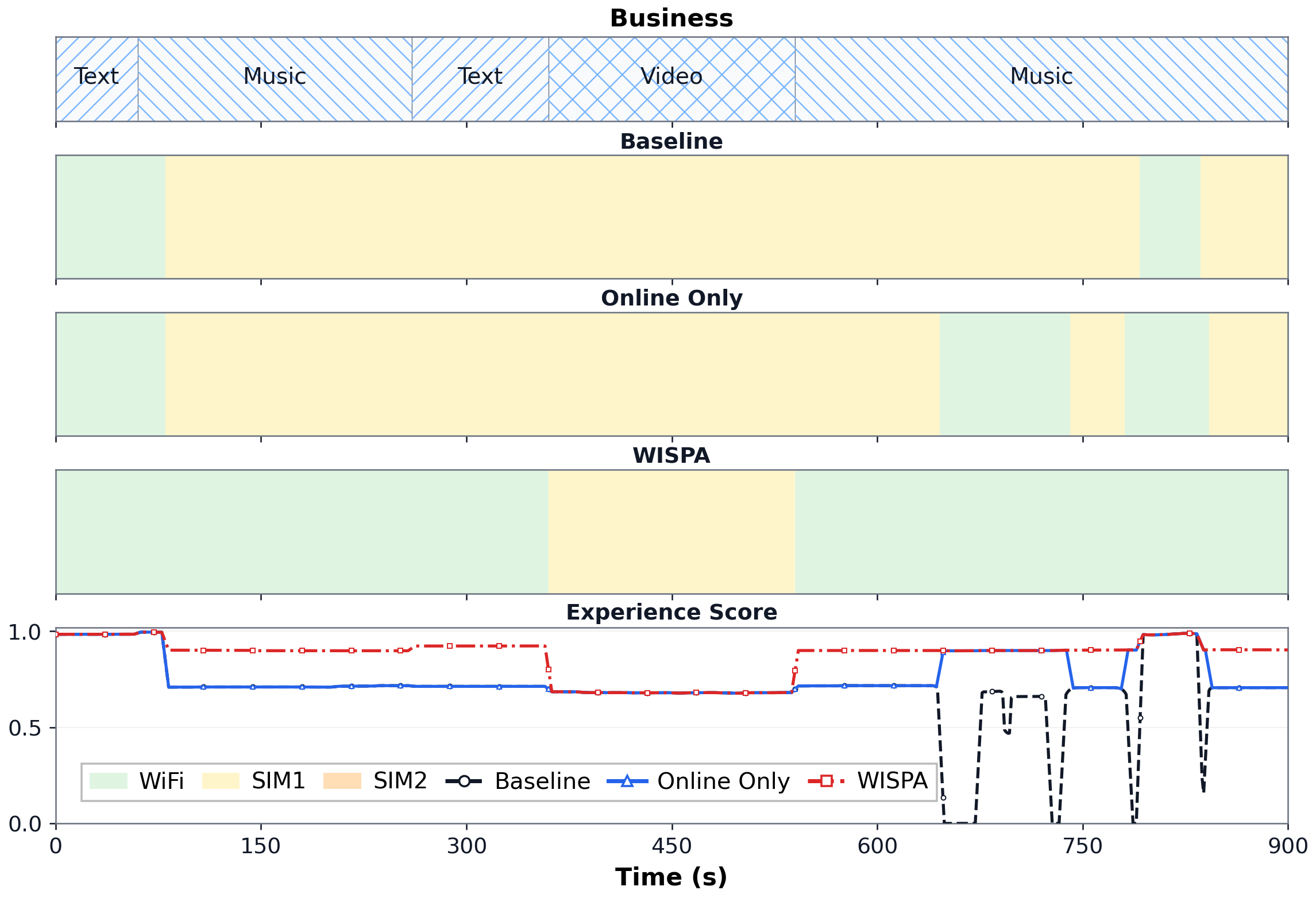}
        \label{fig:shared_cost}}
    \hfill
    \subfloat[Quality Oriented User]{%
        \includegraphics[width=0.49\textwidth]{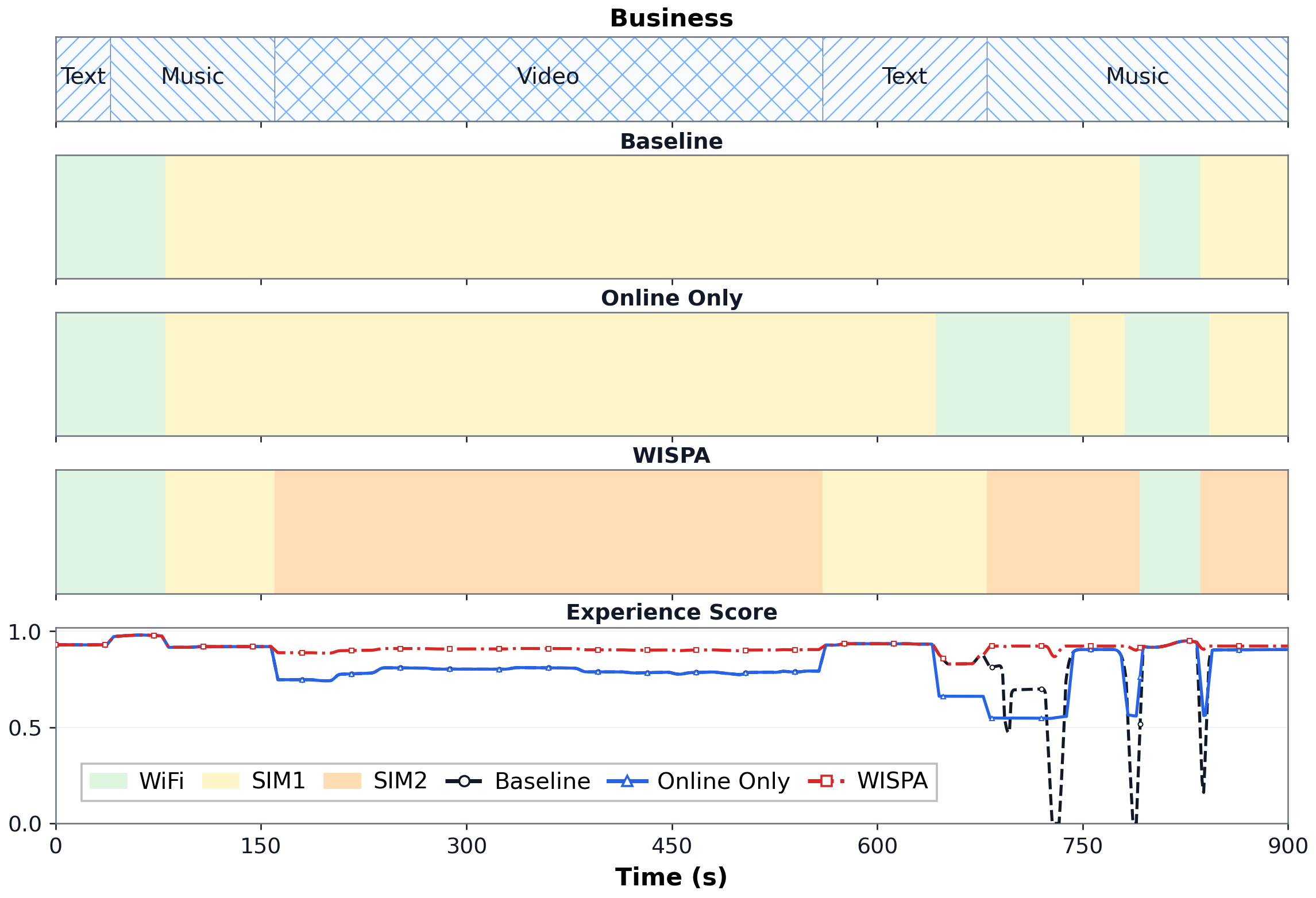}
        \label{fig:shared_quality}}
    \caption{Final strategy trajectories for the Cost Oriented and Quality Oriented users on the shared 900 s campus route. In each subfigure, the top band marks active service windows; the three middle bands show the access strategies of Baseline, Online Only, and WISPA; and the bottom panel reports the corresponding experience-score trajectories.}
    \label{fig:shared}
\end{figure*}
\subsection{Results and Analysis}
\subsubsection{Personalized Heterogeneous Access}
We compare three strategies, i.e., \emph{Baseline}, \emph{Online Only}, and \emph{WISPA}. \emph{Baseline} follows a fixed rule: the terminal uses WiFi in strong-coverage regions and otherwise stays with ordinary cellular access. \emph{Online Only} uses the same online executor as WISPA but keeps the preference parameter at its neutral initial value. \emph{WISPA} starts from the same neutral setting and uses the offline stage to revise the preference parameter from completed usage outcomes. The user profiles represent different preference orientations: the Cost Oriented User favors lower fee burden, the Quality Oriented User prioritizes service assurance, and the Balanced User stays close to the neutral tradeoff. Results are reported using the final strategy obtained after 10 iterations.

Fig.~\ref{fig:bar} first reports the route-level mean experience scores. WISPA improves the score over Online Only from 0.767 to 0.871 (a 13.55\% gain) for the Cost Oriented User and from 0.822 to 0.914 (an 11.25\% gain) for the Quality Oriented User. For the Balanced User, WISPA remains close to Online Only because the neutral initialization already approximates the user's preference, while Baseline stays lower due to its fixed access rule. Since Online Only shares WISPA's deterministic executor and differs only in offline preference calibration, the gains indicate that the offline reflection stage provides useful personalization beyond online scoring alone. Across cycles, the recommended revisions stayed within the admissible bounds, and no invalid recommendation destabilized the loop.

Fig.~\ref{fig:shared} then explains where these gains appear along the shared 900 s campus commute route. For the Cost Oriented User, WISPA raises the experience score by roughly 15--30\% over Online Only in text and music windows where fee-friendly WiFi remains service-acceptable; video calling stays close across strategies because the service floor prevents low-burden access from masking severe service degradation. For the Quality Oriented User, WISPA improves the score by roughly 15--18\% in video calling and later music streaming by using high-assurance access to avoid the drops observed under Online Only and Baseline. The trajectories support the central argument of WISPA: personalization is not a global preference for WiFi or SIM2, but a context-aware adjustment that reinforces low-burden access where acceptable and shifts to high-assurance access where service demand requires it.

\subsubsection{Dynamic Preference Drift}
Personal wireless intelligence should also remain adaptive when user preference changes. We therefore construct a repeated drift experiment in which the preference alternates every 10 usage periods between Cost Oriented and Quality Oriented phases, for 50 periods in total. Strategy state and Memory are preserved across phases, so the agent must revise an existing connection style rather than restart from scratch.

\begin{figure}[!t]
    \centering
    \includegraphics[width=0.48\textwidth]{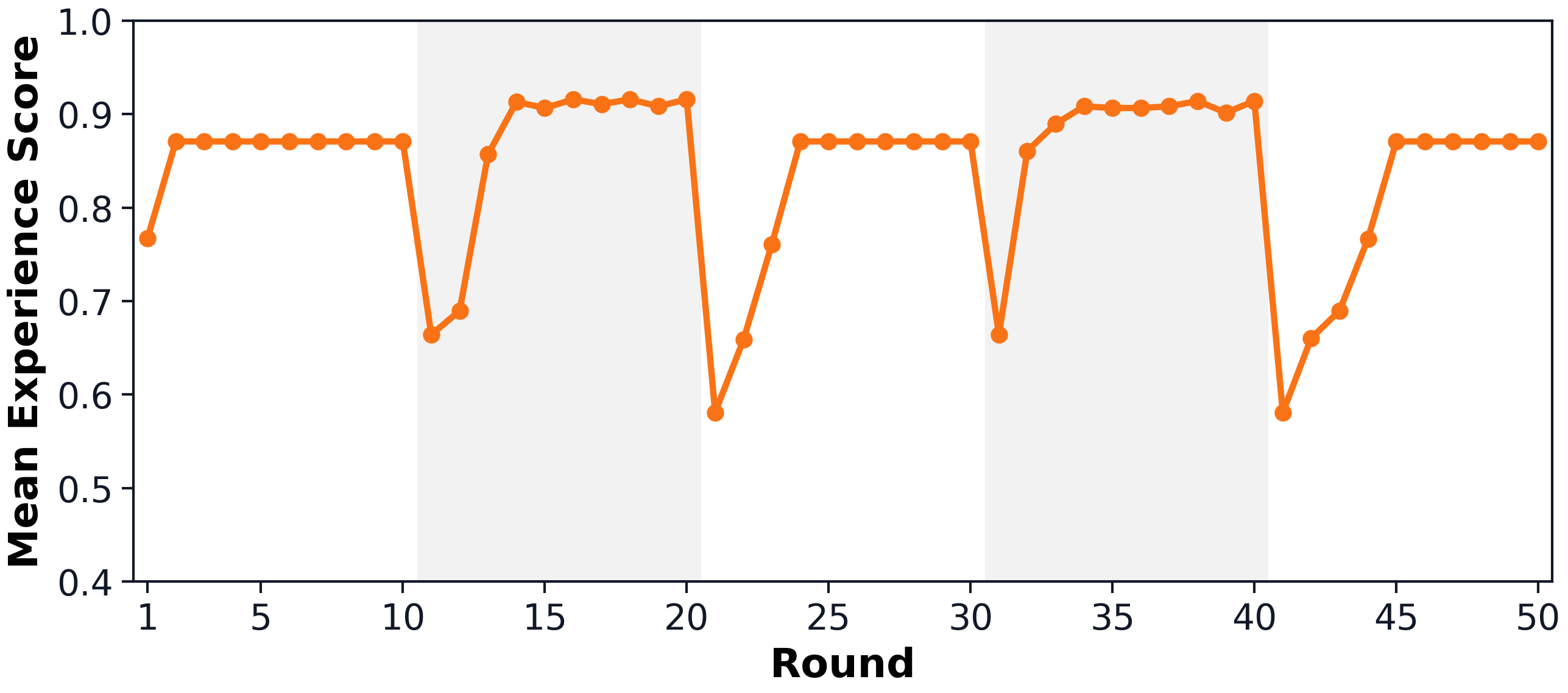}
    \caption{Experience-score evolution under repeated dynamic preference drift. The red curve shows WISPA, and the shaded regions indicate alternating Cost Oriented and Quality Oriented phases.}
    \label{fig:dynamic}
\end{figure}

Fig.~\ref{fig:dynamic} shows the resulting experience-score evolution. Each preference switch creates a temporary mismatch and an immediate score drop, but WISPA recovers within a few periods: it rises above 0.9 in quality-oriented phases and converges to the cost-oriented plateau around 0.87. The repeated drop--recovery pattern indicates that WISPA does not memorize a fixed policy. By preserving Memory across phases, it reuses historical structure while retaining enough plasticity to reshape the terminal's connection style under recurring preference drift. This supports the argument that an offline reflective loop can make terminal-side resource management self-evolving rather than merely preference-fixed.

\section{Discussion and Future Work}
Several issues must be addressed before WISPA-like systems can move from proof of concept to practical deployment. The first is local model execution. Although terminals are becoming more capable, they still operate under strict energy, memory, thermal, and latency constraints, so frequent on-device large-model inference is unrealistic. WISPA therefore confines language-driven reasoning to offline idle windows and keeps the online loop as a lightweight deterministic selector. Future systems should further reduce this burden through compact models, adaptive invocation, and split terminal--edge inference for reflective analysis.

The second issue is security and trustworthiness. A personal wireless agent may process sensitive mobility traces, application records, network choices, and preference signals, and it may face prompt injection, tool misuse, adversarial feedback, or malicious access opportunities. The bounded-step update and fallback rule of the Strategy Module provide an initial safeguard, but future designs need privacy-preserving memory, secure tool authentication, access control, auditable decision logs, and trust-aware resource objectives. These mechanisms are essential if the terminal is to act as a reliable bridge between user intent and network resources.

The third issue is the harness framework introduced in Section~II. LLM agents should not directly manipulate wireless interfaces; instead, the role-specialized modules of WISPA should interact with the terminal through verified harnesses that specify accessible data, callable tools, admissible preference-parameter updates, and safety checks. Such harnesses can translate language-level reasoning into structured and bounded commands, support simulation sandboxes and rollback, and improve interoperability with operating systems, modems, applications, and network APIs.

Finally, the present results remain a single-scenario validation. Broader studies across different environments, mobility routines, service mixtures, access technologies, and user populations are needed, together with comparisons against learning-based baselines. Future research should therefore move from isolated personalization demonstrations toward a safe, efficient, and verifiable ecosystem for terminal-side wireless agents.

\section{Conclusion}
This article presented WISPA, a Wireless Intelligent Self-evolving Personal Agent for terminal-side wireless resource management. We argued that the terminal is a natural place to observe application context, device constraints, mobility routines, and user preferences, and to translate them into personalized wireless decisions. WISPA realizes this idea by separating a lightweight deterministic online executor from an offline pipeline of role-specialized LLM modules, so language-driven reasoning is confined to idle windows while real-time decisions remain deployable. Completed usage traces update Memory and revise the preference parameter through bounded, interpretable steps. A case study on a campus commute route showed that WISPA learns user-specific connection styles, improves the mean experience score by up to $13.6\%$ over a preference-agnostic executor, and re-adapts under repeated preference drift. These results support personal wireless intelligence as a practical complement to network-side optimization, bringing wireless intelligence closer to individual users and their terminals. 

\bibliographystyle{ieeetr}
\bibliography{ref}

\end{document}